\documentclass[pra,aps,twocolumn]{revtex4}
\usepackage{graphicx, epsfig, amsmath, amssymb, amsthm}

\newtheorem{theorem}{Theorem}
\newtheorem{lemma}{Lemma}
\newtheorem{step}{Step}
\newtheorem{substep}{Step}[step]

\def\ket#1{\left\vert #1 \right\rangle}
\def\bra#1{\left\langle #1 \right\vert}
\def\braket#1#2{\left\langle #1 \right\vert \left.\! #2 \right\rangle}

\def\refeqn#1{Eq.\ (\ref{Equation::#1})}
\def\refthm#1{Theorem\ \ref{Theorem::#1}}
\def\reflma#1{Lemma\ \ref{Lemma::#1}}

\begin{document}

\title{Quantum Color-Coding Is Better}

\author{Joshua Von Korff$^3$}
\author{Julia Kempe$^{1,2,4}$}
\affiliation{Departments of Chemistry$^1$, Computer
Science$^2$, and Physics$^3$, University of California, Berkeley, CA 94720\\
$^4$CNRS-LRI, UMR 8623, Universit\'e de Paris-Sud, 91405 Orsay, France }

\date{\today}

\begin{abstract}

We describe a quantum scheme to ``color-code'' a set of objects in order to record which one is which.  In the classical case, $N$ distinct colors are required to color-code $N$ objects.  We show that in the quantum case, only $\frac{N}{e}$ distinct ``colors'' are required, where $e \approx 2.71828$.  If the number of colors is less than optimal, the objects may still be correctly distinguished with some success probability less than $1$.  
We show that the success probability of the quantum scheme is better than the corresponding classical one and is information-theoretically optimal.

\end{abstract}

\maketitle

\section{Introduction} \label{Section::Introduction}

We will describe a quantum scheme for ``color-coding'' a set of objects in order to record which one is which.  That is, we want to be able to tell which is the first object, which is the second, and so on, by looking only at the ``color-code'' quantum labels on the objects.

First, we consider a few examples to clarify the problem.  The classical version of color-coding can be stated as follows:  suppose Alice has $N$ identical boxes.  Inside each box, she writes an integer between $1$ and $N$, using no integer twice.  Alice wants to send the boxes to Bob and have him guess which number is in which box.  Alice helps Bob using a classical color-code: she paints a colored dot (say, red or green) on the outside of each box.  Alice cannot control the order in which Bob receives the boxes, or mark the boxes in any other way.  In other words, Alice is sending the boxes through a classical channel that applies an unknown permutation, and Bob is to guess which permutation was applied.  What procedure should Alice and Bob follow to maximize the probability that Bob will guess the permutation correctly?
This problem is an instance of process tomography - reverse engineering an operation (a permutation in this case) by examining its effects on an initial state.

For $N = 2$ with two colors, Alice need only paint a red dot on box  $1$, and a green dot on box $2$.  Bob can then state the correct order with perfect certainty.  In general, Alice needs $N$ distinct colors if Bob is to distinguish $N$ boxes with perfect accuracy.  In the quantum case, however, we will prove that Alice only needs $\frac{N}{e}$ different colors, where $e \approx 2.71828$ is Euler's constant.

We analyze the classical case first.  Alice is to choose an initial color sequence such as $\psi = $ ``Red Red Green.''  The first color in this list corresponds to box $1$, and so on.  Alice may choose $\psi$ either deterministically or randomly, but randomness never helps her (by concavity \cite{Vonkorff:04}).

Given a $\psi$ with $n$ red dots, the number $n$ is conserved by the permuting channel, so Bob can only receive $N \choose n$ distinct messages.  Then the success probability is at most ${N \choose n}/N! \le {N \choose \frac{N}{2}}/N! = (N! / (\frac{N}{2}!)^2) / N! = 1/ (\frac{N}{2}!)^2$.

To achieve this maximum, Alice could label boxes $1, \ldots \frac{N}{2}$ with red dots, and boxes $\frac{N}{2} + 1, \ldots N$ with green dots.  Bob would then have to guess the ordering within the red set and within the green set.

If Alice is allowed to use $d$ different colors instead of just $2$, her optimal strategy is to label an equal number of boxes with each color, and her success probability is $1/{(\frac{N}{d}!)}^d$  (with slight variations if $d$  does not divide $N$).

\section{Quantum Colors on Three Objects} \label{Section::Quantum}

Now let's consider the quantum version.  Instead of labelling boxes with classical colors (red or green), Alice labels them with quantum spins that can point $\ket{\uparrow}$ or $\ket{\downarrow}$.  As a starting example, suppose there are $N = 3$ boxes.

Alice can ``color'' the boxes with any quantum state $\ket{\Psi} \in (\mathbb{C}^2)^{\otimes 3}$, including entangled states.  When Alice initializes the state, the first copy of $(\mathbb{C}^2)$ corresponds to box number $1$, and so on.    As in the classical case, Alice may as well choose $\ket{\Psi}$ deterministically.

Then Bob receives a state $\Gamma(\sigma) \ket{\Psi}$, where $\sigma$ is a random permutation, and $\Gamma(\sigma)$ is the unitary operator that permutes the $3$ spins via $\sigma$.  Bob wants to perform some measurement on this state that allows him to deduce $\sigma$.

We want to know: can Alice improve on the classical 
protocol, perhaps by entangling the $3$ quantum systems?  Remember that in the classical case, the states that Bob can receive all have the same number of red boxes.  This limits the distinguishability of the received states.  But in the quantum case, Alice can use a signal that is in a superposition of several different numbers of red boxes. So the classical limitation may no longer hold. 

In the classical case, the optimal protocol lets Bob guess the correct permutation with probability $\frac{1}{2}$.  To understand the quantum case, we have to consider the irreducible representations of the action of the permutation group $S_3$ on $V = (\mathbb{C}^2)^{\otimes 3}$.  That is, we must divide $V$ as finely as possible into subspaces that are preserved by $\Gamma(\sigma)$ for all $\sigma$. The vector space is $8$ dimensional, and there are $6$ irreducible representations: $4$ one-dimensional and $2$ two-dimensional.  The one-dimensional representations are the spans of the vectors $\ket{\uparrow \uparrow \uparrow}, \ket{\uparrow \uparrow \downarrow} + \ket{\uparrow \downarrow \uparrow} + \ket{\downarrow \uparrow \uparrow}, \ket{\uparrow \downarrow \downarrow} + \ket{\downarrow \uparrow \downarrow} + \ket{\downarrow \downarrow \uparrow}$, and $\ket{\downarrow \downarrow \downarrow}$.  The first two-dimensional representation is spanned by two vectors, $\ket{1, 1}$ and $\ket{1, 2}$.  Define $\ket{1, 1} \equiv \frac{1}{\sqrt{3}} (\ket{\uparrow \downarrow \downarrow} + e^{2 \pi \imath}/3 \ket{\downarrow \uparrow \downarrow} + e^{- 2 \pi \imath}/3 \ket{\downarrow \downarrow \uparrow})$.  Then the coefficients of $\ket{1, 2}$ are the complex conjugates of the coefficients of $\ket{1, 1}$.  The second two-dimensional representation is spanned by $\ket{2, 1}, \ket{2, 2}$, which are like $\ket{1, 1}, \ket{1, 2}$ except that the directions of all spins are flipped.

Now, suppose Alices uses $\ket{\Psi} = \sqrt{1 / 5} \ket{\uparrow \uparrow \uparrow} + \sqrt{2 / 5} \ket{1, 1} + \sqrt{2 / 5} \ket{2, 2}$.  This state has the interesting property that $|\bra{\Psi} \Gamma(\sigma) \ket{\Psi}| = \frac{1}{5}$ for all $\sigma \ne \epsilon$, where $\epsilon$ is the identity permutation.  That is, all permutations $\Gamma(\sigma)\ket{\Psi}$ of the state $\ket{\Psi}$ are nearly distinguishable from each other,
which hints that it may be a useful state for our purposes.

It turns out that the six positive operators $\{\Gamma(\sigma) \ket{\Psi} \bra{\Psi} \Gamma^\dagger (\sigma)\, | \, \sigma \in S_3\}$ define a POVM (a generalized measurement \cite{Nielsen:00,Preskill:98}) that Bob can use to guess the permutation $\sigma$ that was applied to $\ket{\Psi}$.  The success probability is $\frac{5}{6}$, which is better than the classical $\frac{1}{2}$.

\section{Quantum Color Coding Theorem}

In general, Alice has $N$ boxes, and labels them with $d$-state quantum systems. Let $p(N, d)$ be the probability that Bob measures the permutation correctly for the optimal quantum protocol.  We prove:
\begin{theorem}\label{Theorem::MainTheorem}
Let $r$ be a constant, $d=\lfloor r N \rfloor$. \\
1) If $r > \frac{1}{e}$ then $\lim_{N \rightarrow \infty} p(N, d) = 1$. \\ 2) If $r < \frac{1}{e}$ then $p(N, d) \sim \frac{d^N}{N!}$ as $N \rightarrow \infty$
\end{theorem}
In particular we need only $\approx \frac{N}{e}$ quantum colors to order $N$ objects, a distinct improvement over the classical case! If we have less than $\frac{N}{e}$ colors, we still attain the information-theoretic maximal success probability, $(\mbox{ \# channel states})/(\mbox{ \# message states}) = d^N / N!$ \cite{Vonkorff:04}.  

We prove \refthm{MainTheorem} using the following steps:

\begin{enumerate}
\item First we derive the measurement that Bob can make to determine the correct permutation, under some general assumptions.

\item Next, we maximise this  measurement's success probability, and state it in terms of the dimensions and multiplicities of the irreducible representations of the permutation group $S_N$ on $(\mathbb{C}^d)^{\otimes N}$.

\item Finally, we prove that the success probability satisfies \refthm{MainTheorem}.  This requires an in-depth look at the representation theory of the symmetric group.
\end{enumerate}

\begin{step}
Analyze the possibilities for Bob's measurement.
\end{step}

Our techniques to derive Bob's measurement are inspired by \cite{Massar:95, Gisin:99, Massar:00, Bagan:00, Peres:01, Peres:01b, Fiurasek:02}.
Bob's measurement is a POVM described by positive operators $\{E_\sigma\}, \sum_\sigma E_\sigma = Id$.  Here, $\sigma$ indexes the measurement result, which is a permutation.  Bob ``wins'' if he measures the correct $\sigma$.

Then Bob's success probability, given the state $\Gamma(\sigma) \ket{\Psi}$, is the probability that his measurement result corresponds to the operator $E_\sigma$, which is $P(E, \sigma) = \bra{\Psi} \Gamma^\dagger(\sigma) E_\sigma \Gamma(\sigma) \ket{\Psi}$.  Therefore, if Bob is given a random $\sigma$, his average success probability is $P_{\mbox{av}}(E) =  \frac{1}{N!} \sum_\sigma \bra{\Psi} \Gamma(\sigma)^\dagger E_\sigma \Gamma(\sigma) \ket{\Psi}$.

Consider the new measurement operators $E'_{\sigma '} = \frac{1}{N!} \sum_\sigma \Gamma(\sigma)^\dagger E_{\sigma \circ \sigma '} \Gamma(\sigma)$, where $\sigma \circ \sigma '$ refers to group multiplication in $S_N$.  These "symmetrized" operators are still measurement operators, and give the same success probability as before \cite{Vonkorff:04}.  So we may as well assume that Bob uses such operators.

It is also straightforward to prove that the new operators $\{E'_{\sigma}\}$ satisfy the following useful property:
\begin{equation} \label{Equation::POVMconstraint}
\forall  \sigma :\ \ E'_{\sigma } = \Gamma(\sigma)E'_{\epsilon} \Gamma(\sigma)^\dagger 
\end{equation}
 where $\epsilon$ is the identity permutation. This property simplifies our task, because all relevant positive operators can be deduced from $E'_\epsilon$.  From now on, we drop the prime sign and assume that $E_\sigma$ satisfies \refeqn{POVMconstraint}.

The condition $\sum_\sigma E_{\sigma} = Id$ imposes stringent constraints on $E_\epsilon$.  Using \refeqn{POVMconstraint}
\begin{equation}\label{Equation::POsSumToIdentity}
\sum_{\sigma} \Gamma(\sigma) E_\epsilon \Gamma(\sigma)^\dagger = Id.
\end{equation}
To analyze this constraint, we decompose the space $V=(\mathbb{C}^d)^{\otimes N}$ into irreducible subspaces of the unitary matrices $\Gamma(\sigma)$, corresponding to irreducible representations of $S_N$.  Let $V = \bigoplus_{\rho, b} V^{(\rho, b)}$, where $\rho$ indexes the irreducible representation up to equivalence, and $b$ indexes copies of a given $\rho$.  
Let $\Gamma^{(\rho, b)}(\sigma)$ be the projection of $\Gamma(\sigma)$ onto $V^{(\rho, b)}$.  Then   
\begin{eqnarray}
Id &=&\sum_{\sigma \in S_N} \Gamma(\sigma) E_\epsilon \Gamma(\sigma)^\dagger   
\nonumber \\
&=&\sum_{\sigma} (\sum_{\rho, b} \Gamma^{(\rho, b)}(\sigma)) E_\epsilon (\sum_{\rho ', b '} \Gamma^{(\rho ', b ')}(\sigma)^\dagger) 
  \nonumber \\
&=&\sum_{\sigma, \rho, b, \rho ', b '} \Gamma^{(\rho, b)}(\sigma) E_\epsilon \Gamma^{(\rho ', b ')}(\sigma)^\dagger .  \label{Equation::ConstraintOnE}
\end{eqnarray}
Next, we can select a basis within each irreducible space, indexed by $a$.  That is, $\ket{\rho, b, a}$ are basis vectors for $V$, and the operator $\Gamma^{(\rho, b)}(\sigma)$ has matrix elements $(\Gamma^{(\rho, b)}(\sigma))_{a, a '}$.  We choose the bases of $V^{(\rho,b)}$ such that $\forall \sigma, a, a '$ : $(\Gamma^{(\rho, b)}(\sigma))_{a, a '} = (\Gamma^{(\rho , b ')}(\sigma))_{a, a '}$, i.e. equivalent irreducible representations have the same matrix elements. We can analyze \refeqn{ConstraintOnE} adapting the orthogonality relation for matrix representations \cite{Cornwell:84} to our case:
\begin{eqnarray}\label{Equation::OrthogonalityTheoremEqual}
\sum_{\sigma} (\Gamma^{(\rho b)*}(\sigma))_{\alpha \beta} (\Gamma^{(\rho '  b ')}(\sigma))_{\gamma \kappa} &=&0 \quad \quad \mbox{if} \;  \rho \neq \rho '\nonumber \\
\frac{1}{|G|} \sum_{\sigma} (\Gamma^{(\rho b)*}(\sigma))_{\alpha \beta} (\Gamma^{(\rho   b ')}(\sigma))_{\gamma \kappa} &=& \delta_{\gamma \alpha} \delta_{\beta \kappa} / D_\rho,
\end{eqnarray}
where $D_\rho$ is the dimension of representation $\rho$.

At this point, we will assume that $E_\epsilon = \ket{\Phi} \bra{\Phi}$ for some  state $\ket{\Phi}$, not necessarily  normalized.  
(This assumption will turn out to be a good one: such a POVM is sufficient to establish the result of \refthm{MainTheorem}.)  
Now, we want to find the constraints on $\ket{\Phi}$ resulting from \refeqn{ConstraintOnE}.  We can write $\ket{\Phi} = \sum_{ \rho ,b,a} C_{ \rho ,b, a} \ket{ \rho ,b ,a}$.
Applying \refeqn{ConstraintOnE} and \refeqn{OrthogonalityTheoremEqual}, after a lot of algebra (details in \cite{Vonkorff:04}), we obtain the condition
\begin{equation} \label{Equation::OrthoPhiComponents}
\forall \rho, b, b ' : \quad \sum_a C_{ \rho, b, a} C^*_{ \rho , b ', a} = \delta_{b b '} \frac{D_\rho}{N!}
\end{equation}
Hence the projections of $\ket{\Phi}$ into distinct but equivalent irreducible representations $V^{(\rho , b)}$ and $V^{(\rho , b ')}$ must be ``orthogonal'' in the sense of \refeqn{OrthoPhiComponents}. Since the dimension of the irreducible representation $\rho$ is $D_\rho$, this implies that, for a given $\rho$, $b$ can have at most $D_\rho$ different values. If there are more than $D_\rho$ copies of an irreducible representation $\rho$, the projection of $\ket{\Phi}$ on the remaining copies must be $0$. Let $W$ be the space spanned by all the $V^{(\rho ,b)}$ that have non-zero overlap with $\ket{\Phi}$. The $E_\sigma$ only span the space $W$. To make our measurement a POVM on the whole space $V$ we complete it with the operator $E_{V/W}=Id_{V/W}$. 

Following \refeqn{OrthoPhiComponents}, we can write
\begin{equation} \label{Equation::PhiDecomposition}
\ket{\Phi} = \sum_{\rho, b | V^{(\rho, b)} \subseteq W} \sqrt{\frac{D_\rho}{ N!}} \ket{\Phi_{\rho b}}
\end{equation}
where $\ket{\Phi_{\rho b}}$ is the (normalized) component of $\ket{\Phi}$ in the $(b, \rho)$ subrepresentation.  Without loss of generality we can select  $\ket{\Phi_{\rho b}}$ to be a basis state $\ket{\rho,b,a^{\rho , b}}$ inside $V^{(\rho, b)}$, such that  $a^{\rho ,b'}= a^{\rho , b}$ if and only if $b= b'$.

\begin{step}
With this measurement, what is the maximal success probability?
\end{step}
Now that we have specified a POVM $\{E_\sigma\}$, we want a signal state $\ket{\Psi} \in W$ that maximizes the success probability $P_{\mbox{av}}(E) = \frac{1}{N!} \sum_\sigma \bra{\Psi} \Gamma^\dagger(\sigma) E_\sigma \Gamma(\sigma) \ket{\Psi} = \frac{1}{N!} \sum_\sigma \bra{\Psi} E_\epsilon \ket{\Psi}$ (using \refeqn{POVMconstraint}).  Therefore $P(E) = \bra{\Psi} E_\epsilon \ket{\Psi} = |\braket{\Psi}{\Phi}|^2$.  This is maximized by the choice $\ket{\Psi} \propto \ket{\Phi}$, up to normalization.

Since $\ket{\Psi}$ is normalized and $\ket{\Phi}$ in general is not, we can write
\begin{equation}
\nonumber
P(E)  = |\braket{\Phi}{\Psi}|^2 = \braket{\Phi}{\Phi}
      =  \sum_{b, \rho | V^{(b \rho)} \subseteq W} \frac{D_\rho}{N!} 
      =  \frac{\mbox{dim } W}{N!}
\end{equation}
Where the last steps use \refeqn{PhiDecomposition}.  Then what is the maximum possible dimension of $W$?

To begin with, we have $\dim V = \sum_\rho m_\rho D_\rho = d^N$, where $m_\rho$ is the multiplicity of the irreducible representation $\rho$, and $D_\rho$ is its dimension.  If we could set $W = \oplus_{\rho, b} V^{(\rho, b)} = V$, then $\dim V = \dim W = \sum_\rho m_\rho D_\rho$.  However, as discussed above, $W$ can include at most $D_\rho$ copies of any given $\rho$.  So the maximum dimension of $W$ is $\sum_\rho \mbox{min }(m_\rho, D_\rho) D_\rho$.  Therefore the maximum success probability is given by:
\begin{equation} \label{Equation::MaxProbSuccess}
P_{\max} = \frac{1}{N!} \sum_\rho \min (m_\rho, D_\rho) D_\rho
\end{equation}
\begin{step}  
It remains to evaluate the success probability and show that it is information-theoretically optimal.
\end{step}
We need to 
take an in-depth look at the irreducible representations $\rho$ of the  $S_N$, their dimensions $D_\rho$, and their multiplicities $m_\rho$ in $V = (\mathbb{C}^d)^{\otimes N}$.  As is well-known in representation theory, the irreducible representations of $S_N$ are in one-to-one correspondence with partitions of $N$ as an (unordered) sum of integers.
Such partitions are drawn as Young diagrams, which are rows of boxes, where the total number of boxes is $N$, and each row has no more boxes than the row above it.  The rows are ``left-justified'', i.e. the first boxes from each row form a column.

For the case $r > \frac{1}{e}$, $d = \lfloor r N \rfloor$, we want to prove (\refthm{MainTheorem}) that $P_{\max} \approx 1$ for large $N$.  Our proof of this statement will involve the following steps:

\begin{enumerate}

\item If the columns of a Young diagram $\rho$ are short, then $D_\rho < m_\rho$ (where ``short'' has a precise meaning that will be defined in the proof).

\item Define the Plancherel measure $\mu_N$ of a diagram $\rho$ by $\mu_N(\rho) = \frac{D_\rho^2}{N!}$.  Then the sum of the Plancherel measures of all Young diagrams with ``long'' columns 
approaches $0$ for large $N$.

\item $\sum_\rho \mu_N(\rho) = 1$

\item Using \#2 and \#3, we deduce that $\sum_{\rho \in \{\rho_{short}\}} \mu_N(\rho) \approx 1$, where the sum is over all diagrams with short columns.

\item Using \refeqn{MaxProbSuccess} and \#1, we find that $P_{\max} \geq \frac{1}{N!} \sum_{\rho \in \{\rho_{short}\}} D_\rho^2 =  \sum_{\rho \in \{\rho_{short}\}} \mu_N(\rho) \approx 1$.

\end{enumerate}

\begin{substep}
Young diagrams with short columns have $D_\rho < m_\rho$.
\end{substep}

To determine which of $D_\rho, m_\rho$ is smaller, we can calculate the ratio $D_\rho / m_\rho$, given by \cite{Fulton:91}:
\begin{equation} \label{Equation::DimMultRatio}
\frac{D_\rho}{m_\rho} = \frac{N!}{\prod_{i,j} d - i + j}
\end{equation}
where $(i,j)$ are the row and column of the Young tableau.
\begin{lemma} \label{Lemma::ShortColumns}
Given any $A > 0$ and $r > \frac{1}{e}$, the following statement holds for sufficiently large $N$:  If $d > r N$, then any Young diagram of $N$ boxes whose first column is shorter than $A \sqrt{N}$ must have $D_\rho < m_\rho$.
\end{lemma}

\begin{proof}
If the first column of the diagram is shorter than $A \sqrt{N}$, then all columns are shorter than $A \sqrt{N}$.  So in \refeqn{DimMultRatio}, we must have $i < A \sqrt{N}$.  Therefore the denominator is at least $(d - A \sqrt{N})^N 
> (r- \frac{A}{\sqrt{N}})^N N^N$.  For large enough $N$, $r - \frac{A}{\sqrt{N}} > \frac{1}{e}$.  Let's pick an $\epsilon$ such that $\frac{d}{N} - \frac{A}{\sqrt{N}} > \frac{1}{e}(1 + \epsilon)$.  Then the denominator of \refeqn{DimMultRatio} is greater than $(\frac{N}{e})^N (1 + \epsilon)^N$.  Now, using Stirling's approximation, $N! \approx (\frac{N}{e})^N \sqrt{2 \pi N}$, which is less than $(\frac{N}{e})^N (1 + \epsilon)^N$ for sufficiently large $N$.  Therefore the numerator is smaller than the denominator, and hence $D_\rho < m_\rho$ for sufficiently large $N$.
\end{proof}
\begin{substep}
The Young diagrams with long columns have small total Plancherel measure.
\end{substep}
Let $\lambda_1$ be the length of the first column.  Then \cite{Kerov:03}:
\begin{equation}
\mu_N(\rho) \le e^{- 2 \lambda_1 (\log \frac{\lambda_1}{\sqrt{N}} - 1)}
\end{equation}
According to a theorem of Erd\"{o}s~\cite{Erdos:41}, the total number of diagrams of size $N$ is less than $e^{C \sqrt{N}}$, for some constant $C$. Therefore the total Plancherel measure of diagrams with $\lambda_1 \ge A \sqrt{N}$ is at most $e^{C\sqrt{N}} e^{-2 A \sqrt{N} (\log A - 1)}=e^{(C - 2 A (\log A - 1)) \sqrt{N}}$.  If we choose some $A_0$ such that $ 2 A_0 (\log A_0 - 1) > C$, then this quantity goes to zero as $N \rightarrow \infty$.  Therefore $\lim_{N \rightarrow \infty} \sum_{(\rho | \lambda_1 \ge A_0 \sqrt{N})} \frac{D_\rho^2}{N!} = 0$.
\begin{substep}
$\sum_\rho \mu_N(\rho) = 1$.
\end{substep}
It is well known that $\sum_\rho D_\rho^2 = |G|$, if the sum is taken over all irreducible representations $\rho$ of any group, and $|G|$ is the order of the group.  In our case $|S_N| = N!$, so $\sum_\rho D_\rho^2 / N! = 1$.
\begin{substep}
$\sum_{\rho \in \rho_{short}} \mu_N(\rho) \approx 1$, where the sum is over all diagrams with short columns.
\end{substep}
Combining the results of the two previous steps, we obtain $\lim_{N \rightarrow \infty} \sum_{(\rho | \lambda_1 < A_0 \sqrt{N})} D_\rho^2 / N! = 1$.
\begin{substep} $P_{max} \stackrel{ N \rightarrow \infty} {\longrightarrow} 1$.
\end{substep}
For large enough $N$, all Young diagrams with $\lambda_1 < A_0 \sqrt{N}$ will also have $D_\rho \leq m_\rho$, and
\begin{eqnarray*}
P_{\max} 
& \geq & \sum_{\rho | D_\rho \leq m_\rho} \frac{D_\rho^2}{N!} 
\geq  \sum_{\rho | \lambda_1 < A_0 \sqrt{N}} \frac{D_\rho^2}{N!} \stackrel{ N \rightarrow \infty} {\longrightarrow} 1
\end{eqnarray*}
We have proved \refthm{MainTheorem} for $d > \frac{N}{e}$, so it remains to consider $d < \frac{N}{e}$.  The proof is similar, so we merely hint at which steps are different. First, we replace \reflma{ShortColumns} with:
\begin{lemma} \label{Lemma::ShortRows}
Given any $A > 0$ and $r < \frac{1}{e}$, the following statement holds for sufficiently large $N$: If $d <  r N$, then any Young diagram of $N$ boxes whose first row is shorter than $A \sqrt{N}$ must have $m_\rho < D_\rho$.
\end{lemma}
The proof is almost identical to that of \reflma{ShortColumns}.
Now, the Young diagrams with long rows have small total $\mu_{N, m}$ measure, where each diagram is given weight $\mu_{N,m}(\rho)=\frac{m_\rho D_\rho}{N!}$.
\cite{Kerov:03} states that for large enough $N$, if $\frac{d}{N} \rightarrow r$, we have
\begin{equation}
\mu_{N, d}(\rho) \le e^{- \tilde{\lambda}_1 (2 (\log \frac{\tilde{\lambda}_1}{\sqrt{N}} - 1) - \frac{1}{2 r})}
\end{equation}
where $\tilde{\lambda}_1$ is the length of the first row.
The rest of the proof is completely analogous to the preceding proof.
We find that the success probability is lower bounded by $\sum_{(\rho | \tilde{\lambda}_1 < \tilde{A}_0 \sqrt{N})} \frac{m_\rho D_\rho}{N!} \sim \frac{d^N}{N!}$ as $N \rightarrow \infty$.

Hence we have attained the information-theoretic lower bound for both $r < \frac{1}{e}$ and $r > \frac{1}{e}$, as promised.\\

\section{Concluding notes}\label{Section::Conclusion}
We have shown how quantum coding can give a distinct advantage in protecting information against a random permutation. Note that our results up to Step 3 are completely general and can be formulated for any group.

For instance, suppose we replace $S_N$ with $SO(3)$, the group of rotations in space; and we replace the $N$ $d$-state systems with $N$ $2$-state spins that can be rotated in space.  Then we obtain the problem of transmitting a reference direction using $N$ quantum spins, which has been studied in \cite{Massar:95, Gisin:99, Massar:00, Bagan:00, Peres:01, Peres:01b, Fiurasek:02}, among other papers.  

We hope that our techniques also provide a toolbox for quantum process tomography of other channels.

We thank J. Fern, A. Harrow and O. Regev for discussions and K.B. Whaley for support. 
This work was sponsored by 
DARPA 
and the 
Air Force Laboratory, 
Air Force Material Command, USAF, 
under Contract No. F30602-01-2-0524, 
NSF 
through Grant No. 0121555, ACI-SI 2003-n24 and RESQ IST-2001-37559.

\bibliographystyle{apsrev}

\begin{thebibliography}{14}
\expandafter\ifx\csname natexlab\endcsname\relax\def\natexlab#1{#1}\fi
\expandafter\ifx\csname bibnamefont\endcsname\relax
  \def\bibnamefont#1{#1}\fi
\expandafter\ifx\csname bibfnamefont\endcsname\relax
  \def\bibfnamefont#1{#1}\fi
\expandafter\ifx\csname citenamefont\endcsname\relax
  \def\citenamefont#1{#1}\fi
\expandafter\ifx\csname url\endcsname\relax
  \def\url#1{\texttt{#1}}\fi
\expandafter\ifx\csname urlprefix\endcsname\relax\def\urlprefix{URL }\fi
\providecommand{\bibinfo}[2]{#2}
\providecommand{\eprint}[2][]{\url{#2}}

\bibitem[{\citenamefont{{von Korff} and Kempe}(2004)}]{Vonkorff:04}
\bibinfo{author}{\bibfnamefont{J.}~\bibnamefont{{von Korff}}} \bibnamefont{and}
  \bibinfo{author}{\bibfnamefont{J.}~\bibnamefont{Kempe}}
  (\bibinfo{year}{2004}), \bibinfo{note}{in preparation}.

\bibitem[{\citenamefont{Nielsen and Chuang}(2000)}]{Nielsen:00}
\bibinfo{author}{\bibfnamefont{M.}~\bibnamefont{Nielsen}} \bibnamefont{and}
  \bibinfo{author}{\bibfnamefont{I.}~\bibnamefont{Chuang}},
  \emph{\bibinfo{title}{Quantum Computation and Quantum Information}}
  (\bibinfo{publisher}{Cambridge University Press}, \bibinfo{year}{2000}),
  \bibinfo{note}{p.90}.

\bibitem[{\citenamefont{Preskill}(1998)}]{Preskill:98}
\bibinfo{author}{\bibfnamefont{J.}~\bibnamefont{Preskill}}
  (\bibinfo{year}{1998}), \bibinfo{note}{{\tt http://www.theory.caltech.
  edu/people/preskill/ph229/}, {C}h. 3, 5}.

\bibitem[{\citenamefont{Massar and Popescu}(1995)}]{Massar:95}
\bibinfo{author}{\bibfnamefont{S.}~\bibnamefont{Massar}} \bibnamefont{and}
  \bibinfo{author}{\bibfnamefont{S.}~\bibnamefont{Popescu}},
  \bibinfo{journal}{Phys. {R}ev. {L}ett.} \textbf{\bibinfo{volume}{74}},
  \bibinfo{pages}{1259} (\bibinfo{year}{1995}).

\bibitem[{\citenamefont{Gisin and Popescu}(1999)}]{Gisin:99}
\bibinfo{author}{\bibfnamefont{N.}~\bibnamefont{Gisin}} \bibnamefont{and}
  \bibinfo{author}{\bibfnamefont{S.}~\bibnamefont{Popescu}},
  \bibinfo{journal}{Phys. {R}ev. {L}ett.} \textbf{\bibinfo{volume}{83}},
  \bibinfo{pages}{432} (\bibinfo{year}{1999}).

\bibitem[{\citenamefont{Massar}(2000)}]{Massar:00}
\bibinfo{author}{\bibfnamefont{S.}~\bibnamefont{Massar}},
  \bibinfo{journal}{Phys. {R}ev. {A}} \textbf{\bibinfo{volume}{62}},
  \bibinfo{pages}{040101} (\bibinfo{year}{2000}).

\bibitem[{\citenamefont{Bagan et~al.}(2000)\citenamefont{Bagan, Baig, Brey, and
  Mu{\~n}oz-Tapia}}]{Bagan:00}
\bibinfo{author}{\bibfnamefont{E.}~\bibnamefont{Bagan}},
  \bibinfo{author}{\bibfnamefont{M.}~\bibnamefont{Baig}},
  \bibinfo{author}{\bibfnamefont{A.}~\bibnamefont{Brey}}, \bibnamefont{and}
  \bibinfo{author}{\bibfnamefont{R.}~\bibnamefont{Mu{\~n}oz-Tapia}},
  \bibinfo{journal}{Phys. {R}ev. {L}ett.} \textbf{\bibinfo{volume}{85}},
  \bibinfo{pages}{5230} (\bibinfo{year}{2000}).

\bibitem[{\citenamefont{Peres and Scudo}(2001{\natexlab{a}})}]{Peres:01}
\bibinfo{author}{\bibfnamefont{A.}~\bibnamefont{Peres}} \bibnamefont{and}
  \bibinfo{author}{\bibfnamefont{P.~F.} \bibnamefont{Scudo}},
  \bibinfo{journal}{Phys. {R}ev. {L}ett.} \textbf{\bibinfo{volume}{86}},
  \bibinfo{pages}{4160} (\bibinfo{year}{2001}{\natexlab{a}}).

\bibitem[{\citenamefont{Peres and Scudo}(2001{\natexlab{b}})}]{Peres:01b}
\bibinfo{author}{\bibfnamefont{A.}~\bibnamefont{Peres}} \bibnamefont{and}
  \bibinfo{author}{\bibfnamefont{P.~F.} \bibnamefont{Scudo}},
  \bibinfo{journal}{Phys. {R}ev. {L}ett.} \textbf{\bibinfo{volume}{87}},
  \bibinfo{pages}{167901} (\bibinfo{year}{2001}{\natexlab{b}}).

\bibitem[{\citenamefont{Fiur{\'a}sek et~al.}(2002)\citenamefont{Fiur{\'a}sek,
  Iblisdir, Massar, and Cerf}}]{Fiurasek:02}
\bibinfo{author}{\bibfnamefont{J.}~\bibnamefont{Fiur{\'a}sek}},
  \bibinfo{author}{\bibfnamefont{S.}~\bibnamefont{Iblisdir}},
  \bibinfo{author}{\bibfnamefont{S.}~\bibnamefont{Massar}}, \bibnamefont{and}
  \bibinfo{author}{\bibfnamefont{N.~J.} \bibnamefont{Cerf}},
  \bibinfo{journal}{Phys. {R}ev. {A}} \textbf{\bibinfo{volume}{65}},
  \bibinfo{pages}{040302} (\bibinfo{year}{2002}).

\bibitem[{\citenamefont{Cornwell}(1984)}]{Cornwell:84}
\bibinfo{author}{\bibfnamefont{J.~F.} \bibnamefont{Cornwell}},
  \emph{\bibinfo{title}{Group Theory In Physics}}, vol.~\bibinfo{volume}{1}
  (\bibinfo{publisher}{Academic Press}, \bibinfo{address}{London},
  \bibinfo{year}{1984}), \bibinfo{note}{{A}pp. C, p. 301}.

\bibitem[{\citenamefont{Fulton and Harris}(1991)}]{Fulton:91}
\bibinfo{author}{\bibfnamefont{W.}~\bibnamefont{Fulton}} \bibnamefont{and}
  \bibinfo{author}{\bibfnamefont{J.}~\bibnamefont{Harris}},
  \emph{\bibinfo{title}{Representation Theory: A First Course}}
  (\bibinfo{publisher}{Springer-Verlag}, \bibinfo{address}{New York, NY},
  \bibinfo{year}{1991}), \bibinfo{note}{p. 78, Note that some of the notation
  in the book is inverted with respect to ours.}

\bibitem[{\citenamefont{Kerov}(2003)}]{Kerov:03}
\bibinfo{author}{\bibfnamefont{S.~V.} \bibnamefont{Kerov}},
  \emph{\bibinfo{title}{Asymptotic Representation Theory of the Symmetric Group
  and its Applications in Analysis}}, vol. \bibinfo{volume}{219}
  (\bibinfo{publisher}{AMS}, \bibinfo{address}{Providence, RI},
  \bibinfo{year}{2003}), \bibinfo{note}{p. 122 - 123}.

\bibitem[{\citenamefont{Erd{\"o}s and Lehner}(1941)}]{Erdos:41}
\bibinfo{author}{\bibfnamefont{P.}~\bibnamefont{Erd{\"o}s}} \bibnamefont{and}
  \bibinfo{author}{\bibfnamefont{J.}~\bibnamefont{Lehner}},
  \bibinfo{journal}{Duke Mathematical Journal} \textbf{\bibinfo{volume}{8}},
  \bibinfo{pages}{335} (\bibinfo{year}{1941}).

\end{thebibliography}

\end{document}